\documentclass[structabstract,oldversion]{aa}
\usepackage{graphicx}
\usepackage{txfonts}
\usepackage{color}

\begin{document}
  \title{The large-scale  disk fraction of brown dwarfs in the Taurus cloud as measured with Spitzer}
   \author{J.-L. Monin
          \inst{1}
           \and 
          S. Guieu\inst{2}
           \and
                   C. Pinte\inst{3}
                 \and L. Rebull\inst{2}
\and
           P. Goldsmith\inst{4}
           \and 
           M. Fukagawa\inst{5}
           \and
                       F. M\'enard\inst{1}
           \and D. Padgett\inst{2}
           \and K. Stappelfeld\inst{6} 
           \and C. McCabe\inst{2}
           \and S. Carey\inst{2}
           \and A. Noriega-Crespo\inst{2}
           \and T. Brooke\inst{2}
           \and T. Huard\inst{7}
           \and S. Terebey\inst{8}
           \and L. Hillenbrand\inst{6}
           \and M. Guedel\inst{9}
          \fnmsep\thanks{Based on observations obtained at the Canada-France-Hawaii Telescope (CFHT) which is operated by the National Research Council of Canada, the Institut National des Sciences de l'Univers of the Centre National de la Recherche Scientifique of France, and the University of Hawaii,  and the {\em Spitzer} Space Telescope.}
          }

   \offprints{Jean-Louis Monin}

   \institute{Laboratoire d'astrophysique de Grenoble, Universit\'e Joseph Fourier, CNRS, BP53, 38041 Grenoble, France.
         \and
             {\em Spitzer} Science Center, Mail Code 220-6
Pasadena, CA 91125 USA
             \and University of Exeter, Stocker Road, Exeter EX4 4QL, UK
             \and Jet Propulsion Laboratory
4800 Oak Grove Drive
Pasadena, California 91109 USA
             \and Nagoya University,Furo-cho, Chikusa-ku, Nagoya, 464-8601, Japan
                          \and Caltech, Pasadena, CA 91125 USA
             \and University of Maryland, 1000 Hilltop Circle
Baltimore, MD 21250, USA
             \and Cal State University at Los Angeles, 5151 State University Drive
Los Angeles, CA 90032-8601, USA
             \and Institute of Astronomy, ETH Zurich, 8093 Zurich, Switzerland
             }
   \date{Received 16 April 2009 / Accepted 25 January 2010}

   \abstract{} 
   {The brown dwarf (BD) formation process has not yet been completely understood. To shed more light on the differences and similarities between star and BD formation processes, we study and compare the disk fraction among both kinds of objects over a large angular region in the Taurus cloud. In addition, we examine the spatial distribution of stars and BD relative to the underlying molecular gas}
{In this paper, we present new and updated  photometry data from the Infrared Array Camera
(IRAC) aboard the {{\em Spitzer Space Telescope}} on 43 BDs in the Taurus cloud, and recalculate of the BD disk fraction in this region. 
We also useed recently available CO mm data  to study the spatial  distribution of stars and BDs relative to the cloud's molecular gas.}
{ We find that the disk fraction among BDs in the Taurus cloud is $41\pm12\%$, a value statistically consistent with the one among TTS  ($58\pm9\%$). We find that BDs in transition from a state where they have a disk to a diskless state are rare, and we study one isolated example of a transitional disk with an inner radius of $\approx 0.1\,AU$ (CFHT BD Tau 12, found via its relatively small mid-IR excess compared
to most members of Taurus that have disks. We find that BDs are statistically found in regions of similar molecular gas surface density to those associated with stars.
Furthermore, we find that the gas column density distribution is almost identical for stellar and substellar objects with and without disks.
}
{}

   \maketitle
%

\section{Introduction}


The brown dwarf (BD\footnote{see Section~\ref{sub:bdsample} for a practical definition of what is referred to as a brown dwarf in this paper.}) formation process has not yet been completely understood. Two main classes of models are currently envisioned to explain the origin of BDs. In the "stellar" formation scenario (eg. Padoan \& Nordlund 2002; 2004), BDs form like stars via gravitational collapse and fragmentation of very low-mass cores, followed by significant disk accretion. In the "ejection" formation scenario (Reipurth \& Clarke 2001), BDs are ejected from their parent core and end up being starved for molecular material. In this model, accretion disks around BDs are radially truncated by stripping during the dynamical ejection (see eg. Bate 2009). There is currently no definitive evidence available to reject one or the other scenario either from objects' global properties or from study of individual objects. Studying the disk proportion and structure around BDs may be a key to determining their formation model. 

On the one hand, Luhman et al. (2005; 2006) have studied {\em Spitzer Space Telescope} (Werner et al. 2004) photometry of young objects in Taurus, and from their {\em Spitzer} $3.6$ to $8\,\mu$m colors, found a BD disk fraction of $\approx 40\%$, a value close to that of stars. Guieu et al. (2007) have studied and modeled  the SEDs of 23 substellar objects in the Taurus cloud 
and find that $48\%$ of BDs have disks, a fraction similar to that for T~Tauri stars (TTS) in the same region. In Chamaeleon~I, Damjanov et al. (2007) find no dependence of disk frequency on stellar mass in the K3-M8 spectral type range, well into the BD regime at the age of the cloud ($\approx 2\,$Myr, Luhman 2004). 
The disk phenomenon extends into the planetary mass regime ($M\lesssim 15\,M_{\rm Jup}$). Circumstellar disks have been detected around BDs within the mass range of giant planets ($<15 M_{Jup}$), by Luhman et al. (2005, 2008a,b);
Scholz \& Jayawardhana (2008).
Concerning individual objects, Bouy et al. (2008) have published an extensive data set on the substellar object 2MASS~J04442713+2512164, from the visible to the radio range, including the first photometric measurement of a BD disk at 3.7\,mm, and allowing a detailed analysis of the disk properties. Their analysis shows that this BD has all the characteristics of a star with a disk. 

On the other hand, Luhman et al. (2008a), used {{\em Spitzer}} photometry on a much larger number of objects than Damjanov et al. (2007) in Cha~I, to find a dependence on stellar mass in the disk fraction. 
In Upper Scorpius, Scholz et al. (2007) find a disk frequency  of 37\%+/-9\%, a value higher than what has previously been derived for K0-M5 stars in the same region (at a $1.8\,\sigma$ confidence level), suggesting a mass-dependent disk frequency. \\
Luhman et al. (2007) have presented a disk model for the BD
2MASS J04381486+2611399 showing that the disk has an inner hole and
has an outer radius  ($R_{out} \approx 20-40\,$AU), a size similar to
the predictions from hydrodynamical models ($R_{out} \lesssim 10-40\,$AU) by Bate et al. 2003  and Bate 2009. \\
Thies and Kroupa (2007), studying recent data on the multiplicity properties of stars and BDs show them to have different binary distribution functions. They uncovered a discontinuity of the multiplicity-corrected mass distribution in the very low mass star (VLMS) and BD mass regime, discarding a continous IMF with a high confidence level, and suggesting that VLMS and BDs on the one hand, and stars on the other, are disjoint populations with different dynamical histories.\\ 
In the Taurus cloud, Guieu et al. (2007) find that BD disk fraction appears to vary with the object position within the Taurus filaments, unlike the known young star population.


Disk fraction is therefore one of the keys of the very low mass objects formation models. 
With this purpose in mind, we present in this paper new data about BD disks in the Taurus cloud.
We obtained new or updated {\em Spitzer} Infrared Array Camera (IRAC; Fazio et al. 2004) photometry data on 43 BDs in the Taurus cloud (Section~\ref{sec:obs} and \ref{sec:disk-frac}) and provide an update on the BD disk fraction in this region, complementing the study started by Guieu et al. (2007). We compare the BD disk fraction to the TTS disk fraction in the same region. We present a complete study of the only transition disk found among our BD sample (section~\ref{sec:cfht12}). We also make use of recent CO data (Goldsmith et al. 2008) to study the distribution of stellar and substellar objects relative to the molecular gas (Section~\ref{sec:*bd/gas}) and the distribution of objects with and without disks (section~\ref{sec:obj-dnd}). 


\section{Observations}
\label{sec:obs}
\subsection{Brown dwarf and stellar sample}
\label{sub:bdsample}
	     Our BD sample results from a combination of recent
        studies dedicated to the search for BDs in the Taurus
        cloud. They include studies from Briceno et al. (2002),  Luhman
        (2004), Martin et al. (2002), Guieu et al. (2006), Luhman
        et al. (2006), and Luhman (2006).  All BDs in our sample have been
        spectroscopically confirmed to be late type  ($\geq$
        M6) and members of the Taurus cloud from their low surface
        gravity indicators and/or accretion lines. The stellar-substellar boundary near the spectral type M6 
        at the age of 1-3\,Myr 
        comes from the theoretical models from Baraffe et al. (1998) and Chabrier et al. (2000) and
        the current temperature / spectral type
        scale available for young low mass objects (Luhman et al. 2003).
        The BD surveys 
        used optical and near infrared 2MASS $JHK_s$ photometry. The completeness of these 
        surveys is limited by by the requirement of having a 2MASS counterpart, and should reach $20-30\,M_{\rm jup}$ for $A_V=4$ (Guieu et al. 2006; Luhman 2004). 
	Two objects in the sample (J04335245+2612548 and
        J04242090+2630511) have been found by their infrared excess (Luhman et al. 2006)
         using the Taurus IRAC data; we report their
        photometry in table~\ref{tab:bdphot} but we do not take them into
        account in the computation of the fraction of BDs with disks so that the sample is unbiased in terms of disks. Hence the final number of substellar members used in this study is 41. 
                
Our stellar sample includes all known Taurus members 
in the survey region at the time of this writing.

\subsection{{\em Spitzer} photometry}

Mid-infrared photometry has been obtained with the IRAC instrument at 3.6, 4.5, 5.8, and $8\,\mu$m 
on board the {\em Spitzer Space Telescope}. 
The {\em Spitzer} fluxes were extracted from the mosaic images obtained as
part of  General Observer programs 3584, 30816 and 462 for the  mapping of the Taurus cloud (Padgett et al. 2007; Padgett et al., in prep.).
The observations were carried out during 2005 and 2007, with a 2$\times$12 second high dynamic range frame per epoch for IRAC (two 0.4~sec exposures and two 10.4~sec exposures). 

Data reduction and source extraction are described in detail in the Taurus data delivery document (Padgett et al. 2008) and in Padgett et al. (in prep.). A comprehensive study of the stellar population
in Taurus using the Spitzer Space Telescope can be found in Rebull et al. (2009).
Briefly, the IRAC basic calibrated data (BCDs) have been processed by the IRAC artifact mitigation software (available on the SSC website) before being assembled in mosaics. Aperture photometry has been performed on the source detection output of the APEX detection algorithm. Fluxes have been extracted in an aperture of 2 pixel radius. The sky has been measured in 2 to 6 pixel radii. Fluxes were aperture corrected using the corrections listed in the IRAC Data Handbook. They have been converted to magnitudes given the zero magnitude fluxes of 280.9\,$\pm$4.1, 179.7\,$\pm$2.6, 115.0\,$\pm$1.7
and 64.1\,$\pm$0.9 mJy for channels 1, 2, 3, and 4 respectively (Reach et al. 2005). 
The typical magnitude error is 0.05 for sources brighter than 10, 10, 8.5, 8 mag for channels 1, 2, 3 and 4 respectively, and increases to 0.07, 0.07, 0.10 \& 0.10 for the faintest sources at 14, 13.5, 13.5, 12.7 mag. 


Guieu et al. (2007) have published {\em Spitzer} photometry for 23 BD in the Taurus cloud based on the original {\em Spitzer} Taurus survey. Since then, the {\em Spitzer} survey has been extended to a larger region (Padgett et al. 2009) and new data are now available for more BDs. 
In Table~\ref{tab:bdphot}, we list 43 BDs currently known in the Taurus cloud  with {\em Spitzer} IRAC  photometry. The data from Guieu et al. (2007) have been (slightly) updated and are therefore superseded by these new values. 
Among the 20 additional objects reported here, listed in table~\ref{tab:bdphot}, 9 show an infrared excess, taken as a signature of a circumstellar disk (see section~\ref{sec:disk-frac}). The table also includes visual absorption values (col 10). These values come from the literature when available, they are often converted from $A_J$ to $A_V$ using $A_J=0.265\,A_V$, or estimated from their spectral types combined with their J-H color

\begin{table*}[!htb]
\caption { IRAC {\em Spitzer} photometry for BDs. $^a$ denotes the two objects found from their IR excess. \label{tab:bdphot}}
\begin{tabular}{llllllllll} \hline\hline
Name         & 2MASS name &  N(H$_2$)                          & Sp Type  & Disk &      [3.6] &    [4.5]  &   [5.8]  &   [8]  &  A$_{\rm V}$ \\
    	          &                         &  ($10^{21}\,$cm$^{-2}$)     &               &       presence      &         &          &        &                  &                     \\ 
J041411+281153	&   04141188+2811535    & 5.89  &   M6.25            &  yes  & 10.93  &  10.40   & 10.12 &      8.99       &     1.06  \\
KPNO-Tau~1	     & 04151471+2800096        & 3.60  &   M9.00            &  no  & 13. 23 & 12.72   &   12.94          & 	   12.81           &     0.41 \\
 KPNO-Tau~2      & 04185115+2814332         & 5.88   &   M6.75          & no   & 12.25   &  12.11  &  12.02    &  11.84 &     0.37 \\ 
 KPNO-Tau~12   & 04190126+2802487         & 4.98   &   M9.00          &  yes  & 13.97   &  13.61   &  13.23   &  12.75  &     0.53 \\ 
 CFHT-Tau~14     & 04221644+2549118        & 1.86   &   M7.75          & no   & 11.48  &  11.34    &  11.28   &  11.23  &  0.57 \\ 
 CFHT-Tau~9        & 04242646+2649503      & 2.04   &   M6.25          &yes    & 11.16   &  10.88   &  10.51   &  9.83    &  0.91 \\ 
 J042728+261205  &   04272799+2612052    & 2.03   &   M9.50          & no  &  12.57   &  12.37   &  12.21   &  12.08  &   2.45 \\ 
 CFHT-Tau~15    & 04274538+2357243        & 1.88   &   M8.25          &no  &   13.24  &  13.15    &  13.25   &  13.06  & 1.30 \\ 
 KPNO-Tau~5      & 04294568+2630468       & 1.92   &   M7.50          & no  &  11.05  & 11.02    & 10.94    & 10.83   &     0.00 \\ 
 KPNO-Tau~6    & 04300724+2608207          & 2.39   &   M9.00          & yes  &  13.12  & 12.77    & 12.42   & 11.58     &   0.88 \\ 
 CFHT-Tau~16    & 04302365+2359129        & 2.42   &   M8.50          &  no  & 13.23  & 13.16   & 13.04   & 12.99    &   1.51 \\ 
 KPNO-Tau~7    & 04305718+2556394          & 2.25   &   M8.25          &yes  &   12.62  & 12.29   & 11.99   & 11.25    &  0.00 \\ 
 CFHT-Tau~13    & 04312669+2703188          & 3.83   &   M7.25          &  no  &  12.90  & 12.75  & 12.72   & 12.67   &    3.49 \\ 
 CFHT-Tau~7      & 04321786+2422149          & 5.43  &   M6.50          &  no  &   9.98 &  9.87 &  9.76        &  9.72 &    0.00 \\ 
 CFHT-Tau~5      &    04325026+2422115     & 6.10  &   M7.50          &  no &    10.46  & 10.27  & 10.10  & 10.07 &    9.22 \\ 
 CFHT-Tau~12    & 04330945+2246487         & 3.72   &   M6.50          &  yes  &  10.86  & 10.63  & 10.34   &  9.95   &   3.44 \\ 
 CFHT-Tau~1   & 04341527+2250309           & 3.56   &   M7.00          & no  &    11.24  & 11.10  & 10.98  & 11.02 &    3.10 \\ 
 CFHT-Tau~11   & 04350850+2311398         & 2.16   &   M6.75          &  no &    11.19  & 11.12  & 11.04  & 10.99 &   0.00 \\ 
 KPNO-Tau~9     & 04355143+2249119          & 2.03   &   M8.50          & no &     13.63  & 13.52  & 13.67  & 13.41 &      0.00 \\ 
 CFHT-Tau~2    & 04361038+2259560            & 3.07   &   M7.50          &  no &     11.63  & 11.43  & 11.34  & 11.33 &   0.00 \\ 
 CFHT-Tau~3     & 04363893+2258119         & 2.55   &   M7.75          &  no  &    11.79  & 11.69  & 11.60  & 11.57 &   0.00 \\ 
 J043801+255857  & 04380083+2558572   & 4.40  &   M7.25          &   no  &     9.60  &  9.47  &  9.37  &  9.31  &   0.64 \\ 
 J043814+261139  & 04381486+2611399   & 4.20   &   M7.25          & yes &       10.80  & 10.21  &  9.64  &  8.92 &    0.00 \\ 
 GM~Tau             &     04382134+2609137     & 5.24   &   M6.50          & yes &        9.27   & 8.77   & 8.43  &  7.81 &    4.34 \\ 
 J043904+254426   &  04390396+2544264  &  4.88  &   M7.25          & yes &       10.75  & 10.46  & 10.02  &  9.14 &    0.40 \\ 
 CFHT-Tau~4     & 04394748+2601407          & 6.21   &   M7.00          &  yes&        9.54   & 9.07  &  8.60  & 7.79  &    2.64 \\ 
 J044111+255511   &  04411078+2555116   & 5.07   &   M6.50   &  yes        &      10.78  & 10.35 &   9.92  & 9.22 &     1.80 \\ 
 J044148+253430   &    04414825+2534304  & 5.84   &   M7.75    & yes  &             11.43  & 10.93  & 10.50   & 9.54 &     1.06 \\ 
 J044427+251216   & 04442713+2512164  & 3.10   &   M7.25   &  yes&               9.56   & 9.00 &  8.36 &  7.43  &    0.00 \\ 
 J041524+291043 & 04152409+2910434 & 2.26   &   M7.00   &  no  &            11.86  & 11.79  & 11.66  & 11.49 &    1.18 \\ 
 J041618+275215  & 04161885+2752155 & 4.83  &  M6.25  &   no &             10.88  & 10.78  & 10.67  & 10.68  &   1.07 \\ 
 J042900+275503 &  04290068+2755033 & 1.99  &   M8.25  &   yes  &            12.30  & 11.99  & 11.60  & 10.92  &    0.00 \\ 
 J043119+233504  & 04311907+2335047& 2.27  &   M7.75   &  no  &            11.66  & 11.53  & 11.56  & 11.46  &    0.90 \\ 
 J043203+252807  & 04320329+2528078 & 1.91  &   M6.25   &  no  &            10.30  & 10.20  & 10.13  & 10.09 &    0.00 \\ 
 J043223+240301  & 04322329+2403013 & 3.01  &   M7.75   &  no  &            10.89  & 10.83  & 10.79  & 10.68 &    0.00 \\ 
 J043342+252647  & 04334291+2526470& 2.71  &   M8.75   &  no  &            12.76  & 12.63  & 12.52  & 12.47  &    0.62 \\ 
 J043545+273713 & 04354526+2737130 & 1.65  &   M9.25   &  no  &            13.18 & 13.11  & 12.93  & 13.07  &       0.28 \\ 
 J043610+215936  & 04361030+2159364 & 1.08  &   M8.50   &  yes &             13.02  & 12.74  & 12.41  & 11.74  &  0.33 \\ 
 J042420+263051$^a$  &  04242090+2630511 & 3.77  &   M6.50   &  yes  &            11.83  & 11.45  & 11.00  & 10.40   &   0.85 \\ 
 J043352+261254$^a$  &  04335245+2612548& 3.78  &   M8.50   &  yes  &             13.17 & 12.66  & 12.28  & 11.40  &    5.20 \\ 
 J042630+244355 & 04263055+2443558 &  3.37   &  M8.75    &   yes   &          12.57  & 12.21  & 11.76  & 11.08  &     0.00  \\
 J042154+265231 & 04215450+2652315 &  5.30  &  M8.50    &   no   &            13.22 &  13.12  & 12.90 & 12.80    &       2.97  \\
CFHT-Tau~10	    & 04214631+2659296       &   3.53  &  M6.25   &   yes  &             11.54 &  11.32 & 11.05	  & 10.45   &    3.59 \\
 \hline
   \hline
\end{tabular}
\end{table*}

\subsection{CO data}
In order to study the distribution of BD with and without disks with respect to the surrounding molecular cloud, we use the recently published CO data from Goldsmith et al. (2008). They have used $^{12}$CO and  $^{13}$CO data to compute the molecular hydrogen column density, N(H$_2$), over a large area of the Taurus cloud, encompassing all the BD of our sample. The distribution of N(H$_2$) at the position of every BD of the sample is listed in table~\ref{tab:bdphot} (see section~\ref{sec:*bd/gas} for a description of how the density is computed for every object). 

\section{Brown dwarf disks}
\subsection{Update of the brown dwarf disk fraction}
\label{sec:disk-frac}
\begin{figure}[!htb]
\includegraphics[width=\hsize]{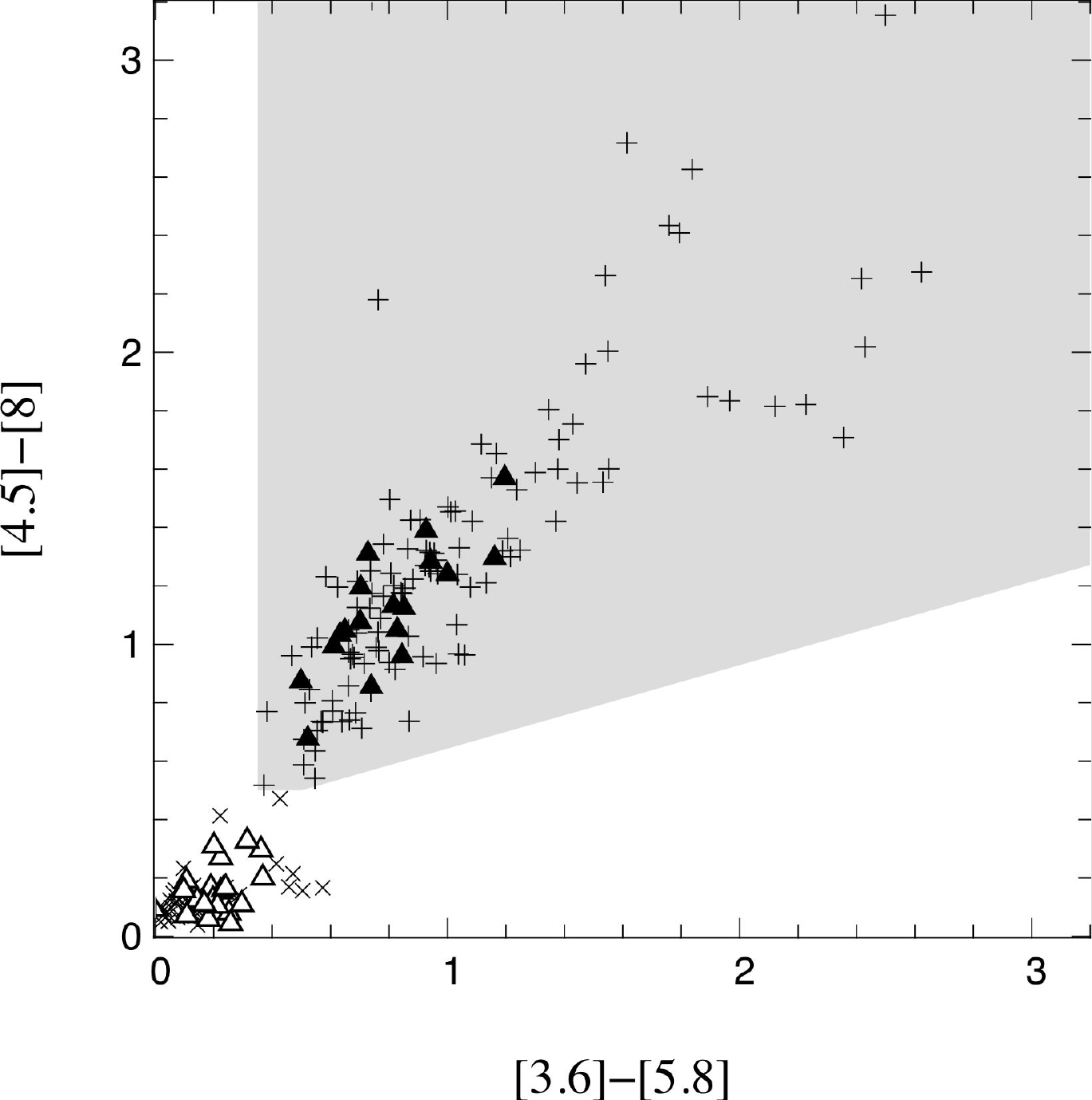}
\caption{Taurus TTS and BDs in a {\em Spitzer} IRAC color-color diagram. Stars without disks are plotted as '$\times$' symbols while stars with disks are plotted as  '$+$'  symbols. BDs are plotted as empty or solid triangles depending on  their colors / disk status. \label{fig:colcol}}
\end{figure}

We use the {\em Spitzer} color information to decide whether a BD (or a TTS) has a disk or not. Following Gutermuth et al. (2008), all Taurus members with the following color constraints are likely BDs or stars with disk.
\begin{equation}
  \left\{\begin{array}{c}
      \left[4.5\right]-[8]>0.5\\
      \&\\
      \left[3.6\right]-[5.8]>0.35\\
      \&\\
      \left[3.6\right]-[5.8]\le \frac{0.14}{0.04} \times \left([4.5]-[8]-
        0.5\right)+0.5 
    \end{array}\right.
\end{equation}
%
%
In fig~\ref{fig:colcol}, we have plotted the [4.5]$-$[8] vs. [3.6]$-$[5.8] {\em Spitzer} color-color data for all the BD from table~\ref{tab:bdphot} (triangles) superimposed on the same plot for previously known Taurus stars with ($\times$ symbols) and without ($+$ symbols) disks. The grey area delineates the objects with a disk. Apart from some 
%
%
objects at the edge of this zone, 
objects (TTS and BDs) with and without infrared excess are well separated in the plot. 
%

Almost all the BDs in table~\ref{tab:bdphot} have $A_V \lesssim4$ (one has $A_V=5.2$ and one has $A_V=9.2$). We choose to restrict our analysis to the part of the cloud with $A_V\le4$, for BDs and for stars. Removing the two objects found from their disk excess and mentioned in section~\ref{sub:bdsample}, and two more from this absorption constraint yields a BD disk fraction of $16/39=41\pm 12\,\%$.
If we use the {\em Spitzer} photometry to compute the fraction of disks in the TTS, we find that among 103 stars with $A_V<4$, 60 have disks, yielding a disk fraction of $58\pm9\,\%$. 
The disk fraction of TTS and BD in Taurus thus appears identical at the one sigma level.

\subsection{Brown dwarf transition disks}
\begin{figure}[htb]
\includegraphics[width=\hsize]{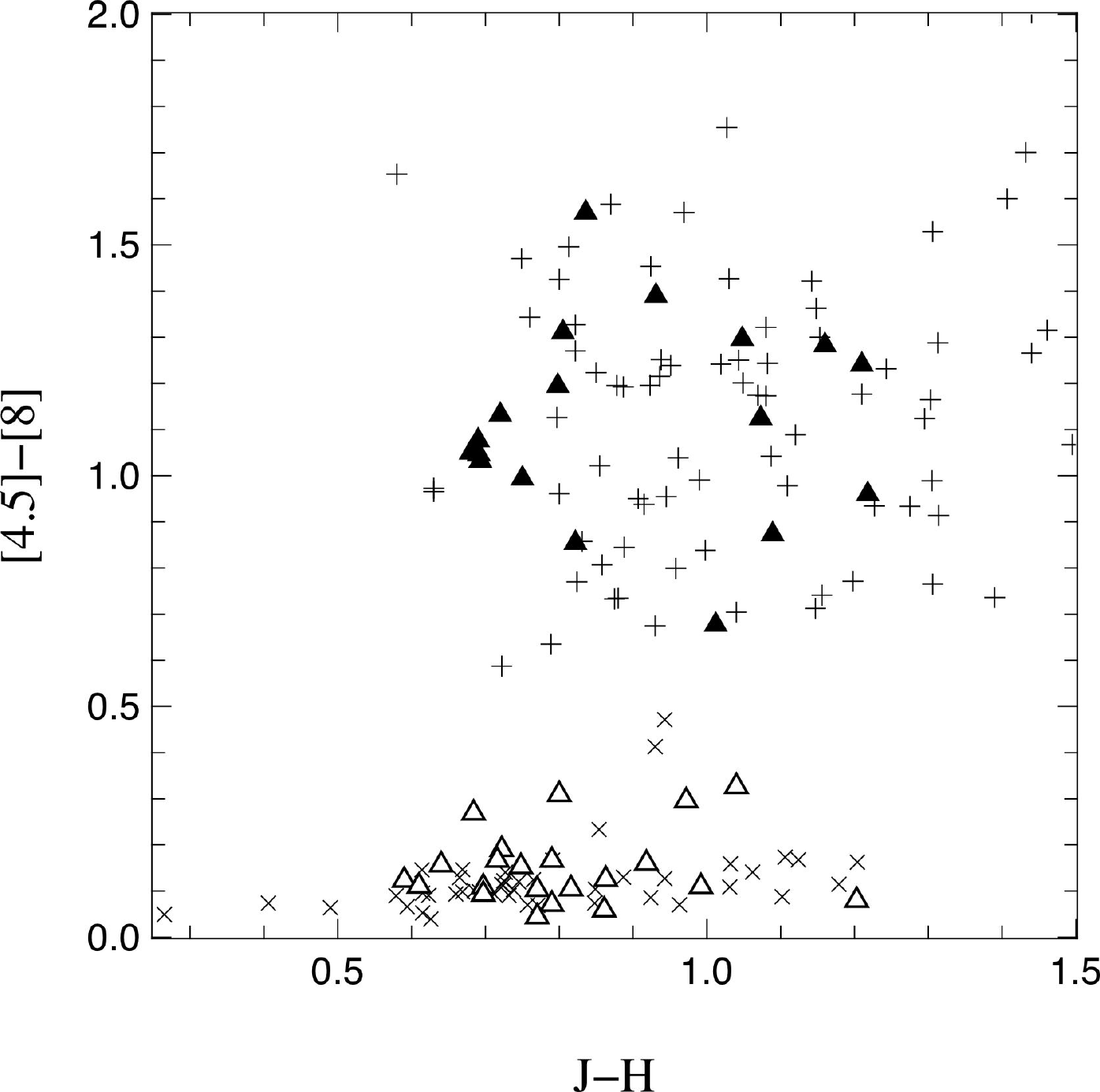}
\caption{Taurus color-color diagram in the $[4.5]-[8] / J-H$ plane. Symbols are identical to those in figure~\ref{fig:colcol}. Note that this figure is focused on the J$-$H BD regime. Twenty more TTS are distributed between $J-H$=2 and $J-H$=3.5\label{fig:i2i4-jh}}
\end{figure}

The plot in figure~\ref{fig:colcol} shows two well separated populations. However, the plot is crowded and makes the reading of possible transition objects difficult. In order to ease the measurement of the color excess, we 
follow  Damjanov et al (2007) in plotting [4.5]$-$[8] vs. J$-$H, this latter color being used as a proxy for the objects photosphere, for all the sources of our sample in figure~\ref{fig:i2i4-jh}.

While the values of J$-$H span a wide range of values due to the distribution of spectral types, the [4.5]$-$[8] colors  fall into two groups, depending on whether or not the objects have disks. The objects with disks show more scatter in their [4.5]$-$[8]
     index since it is dominated by disk excess emission. 
Figure~\ref{fig:i2i4-jh} shows some transition objects and confirms that CFHT-Tau 12 is a BD with a disk of this type. We study this peculiar object in more details in the following section.

\section{CFHT Tau 12}
\label{sec:cfht12}
CFHT~Tau~12 presents a very peculiar SED, with an excess starting at
$8\,\mu$m and extending to longer wavelengths.
In order to ascertain the nature of the circumstellar environment of CFHT~Tau~12,
we have performed a detailed modeling of its SED.

\subsection{Disk model}

We use the 3D Monte-Carlo continuum radiative
transfer code MCFOST (Pinte et al. 2006).
Our model includes multiple scattering,
dust heating assuming radiative equilibrium,
and continuum thermal re-emission.

We use a density distribution with a Gaussian vertical profile
$\rho(r,z)=\rho_0(r)\,\exp(-z^2/2\,h^2(r))$, assuming a
vertically isothermal, hydrostatic, non self-gravitating disk.
Power-law distributions are utilized for the surface density
$\Sigma(r) = \Sigma_0\,(r/r_0)^{\alpha}$  and the scale height $
h(r) = h_0\, (r/r_0)^{\beta}$ where $r$ is the radial coordinate
in the equatorial plane, $h_0$ the scale height at the radius
$r_0 = 100$\,AU. The disk extends from an inner radius  $r_{\rm
in}$  to an outer limit radius $r_{\rm out} = 300$\,AU. The central star
is represented by a uniformly radiating sphere with a NextGen spectrum
from Allard et al. (2000) ($T_\mathrm{eff} = 2\,900$\,K, $\log g$ = 3.5)
and an $A_V = 3.44$.

We consider homogeneous spherical grains, using the
dielectric constants described by Mathis \& Whiffen (1989) in
their model A. The
differential grain size distribution is given by $\mathrm{d}n(a)
\propto a^{-3.5}\,\mathrm{d}a $ with grain sizes between
$a_{\mathrm{min}} = 0.03 \mu$m and  $a_{\mathrm{max}}$, which is taken
as a free parameter.
The mean grain density is $0.5$ g.cm$^{-3}$ to account for
fluffiness. Extinction and scattering opacities, scattering
phase functions and Mueller matrices are calculated using Mie
theory. Dust and gas are assumed to be perfectly mixed, and grain
properties are taken to be independent of position within the
disk. 

\subsection{Model fitting}
\label{sec:analysis}

Due to the ambiguities with SED fitting, it is not
possible to constrain model parameters independently. The
robust estimation for the range of validity of the parameters instead
requires the potential correlations between each of the parameters to
be taken into consideration. With this in mind, we systematically
explored a grid of models by varying 7 free parameters whose values
are listed in Table~\ref{tab:parameter_space}.

\begin{table}[htb]
  \caption{Range of values explored for each
    parameter.\label{tab:parameter_space}}
  \begin{tabular}{lllll}
    Parameter & min value & max value & \# values & sampling\\
    \hline
    $r_\mathrm{in} \ ({\rm AU})$   & 0.01   &  100  & 20  & log    \\
    $M_\mathrm{dust}$ (M$_\odot$)   & $10^{-13}$   &  $10^{-6}$  & 8  & log    \\
    $a_\mathrm{max} \ ({\rm AU})$  & 1   & 1\,000& 10  & log    \\
    $\beta$                     & 1.0 & 1.25  & 5   & linear \\
    $\alpha$                    &-0.5 & -1.5  & 5   & linear \\
    $h_0$                       & 10  &  30   & 10  & linear  \\ 
    $cos(i)$                    & 0.05& 0.95  & 10  & linear  \\ 
    \hline
  \end{tabular}
\end{table}

Comparisons between the models and the observations 
 were drawn according to reduced $\chi^2$ calculations. 
To determine the range of validity for the parameters, we
used a Bayesian inference method
(Press et al. 1992; Lay et al. 1997; Pinte et al. 2007; Pinte et al. 2008). This technique allows us to
estimate the probability of occurrence of each parameter value. The
relative probability of a single point of the parameter space
(\emph{i.e.} one model) is proportional to $\exp (-\chi^2/2)$, where
$\chi^2$ refers to the reduced $\chi^2$ of the corresponding
model. All probabilities are normalized at the end of the procedure so
that the sum of the probabilities of all models over the entire grid
is equal to 1.  


The Bayesian method relies on \emph{a priori} probabilities for the
parameters.  Here, we assume that we do not have any
\emph{preliminary available information}, and we choose uniform a priori
probabilities which correspond to a uniform sampling of
the parameters. However, in the absence of any data, some grid points are
more likely than others: consideration on solid angles show that an
inclination between 80 and 90$^\circ$ (close to edge-on) is more
likely than a inclination between 0 and 10$^\circ$. Uniformly
  distributed disk inclinations and orientations in three dimensions 
correspond to a uniform distribution in the cosine of the inclination.
Also, some physical quantities,
like the inner radius, tend to be distributed logarithmically. The
grid was built according to these distributions (Table~\ref{tab:parameter_space}).

\begin{figure*}[htb]
         \includegraphics[angle=0,width=\hsize]{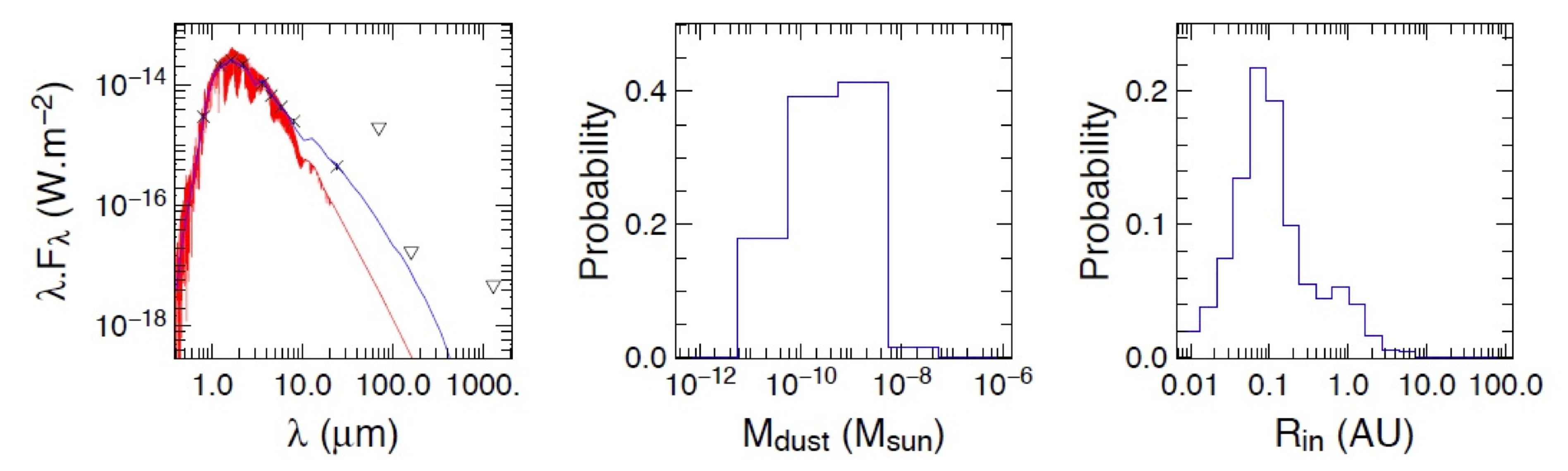}
          \caption{SED modeling of CFHT~Tau~12. \emph{Left:} SED and best
    model. Crosses: data points, triangles: upper limits. Full blue line: synthetic SED. Red
    line: stellar photosphere. \emph{Center:} marginal Bayesian
    probability for the disk mass. \emph{Right:} marginal Bayesian
    probability for the inner radius. 
    \label{fig:cfht_tau12}}
\end{figure*}

\subsection{Results}

Figure~\ref{fig:cfht_tau12} presents the best fit SED (\emph{i.e.} the
model with the smallest $\chi^2$) and the relative figures of merit
estimated from the Bayesian inference method for the disk inner
radius, and for the inner radius and maximum grain size
simultaneously.
 These results were obtained from marginalization (\emph{i.e.}
summing) of the probabilities of all models, where one (or two) parameter is
fixed successively to its different values. The resulting histograms
indicate the probability that a parameter takes a certain value, given
the data and assumptions of our modeling. The width of the
probability curve in the central panel is a strong indicator of how
well the disk mass is constrained. 

The range of validity for the disk mass is defined as the interval
[M$_\mathrm{1}$, M$_\mathrm{2}$] where: 
\begin{equation}
  \int_{M_\mathrm{min}}^{M_\mathrm{1}} p(r)\,\mathrm{d} r =
  \int_{M_\mathrm{2}}^{M_\mathrm{max}} p(r)\,\mathrm{d} r =
  \frac{1-\gamma}{2}  
\end{equation}
with $\gamma = 0.997$. The interval is [$5\,10^{-12}$, $5\,10^{-9}$
M$_\odot$] (Fig~\ref{fig:cfht_tau12}, central panel) and corresponds
to a 99.7\,\% confidence interval, \emph{i.e.} equivalent to a 3\,$\sigma$ interval
in the case of a Gaussian distribution of probability. This constraint
is set by the upper limits at 70 and 160\,$\mu$m.
 Unsurprisingly,
most of the parameters are not, or only slightly, constrained. 
For instance, the same
analysis for the inner radius (Fig~\ref{fig:cfht_tau12}, right
panel) gives an interval of [0.01, 1.5]\,AU.

CFHT~Tau~12 is surrounded by a very low mass disk, with a dust mass
lower than $1.5\,10^-3$ Earth mass. This
is significantly lower than disk masses measured so far for BDs
(Scholtz et al. 2006). The mass we derive is comparable to 
masses of debris disks around M~stars,
which have been estimated to be between $10^{-2}$ and few earth masses (see
for instance Fig.~3 of Wyatt 2008). Because the disk excess is
only detected in the mid-infrared, an hypothesis is that most of the
outer disk has been removed (potentially via an 
ejection mechanism of the BD for instance). 

\section{Spatial distribution of brown dwarfs in the Taurus cloud}
\subsection{Stars and brown dwarfs relative to Taurus molecular gas} 
\label{sec:*bd/gas}
Luhman (2006) studied the nearest neighbor distance and found no differences between the star and BD distributions. Such a result is consistent with recent computations by Bate (2009) who finds no significant mass segregation. 
However, models of spatial distribution of BDs produced by the decay of small-N stellar systems by Goodwin et al. (2005)  show that similar distributions sometimes arise even with ejections. In other words, a different spatial distribution for stars and BDs is probably a signature of ejections, but a lack of difference does not necessarily exclude the ejection scenario.

\begin{figure}[htb]
\includegraphics[width=\hsize]{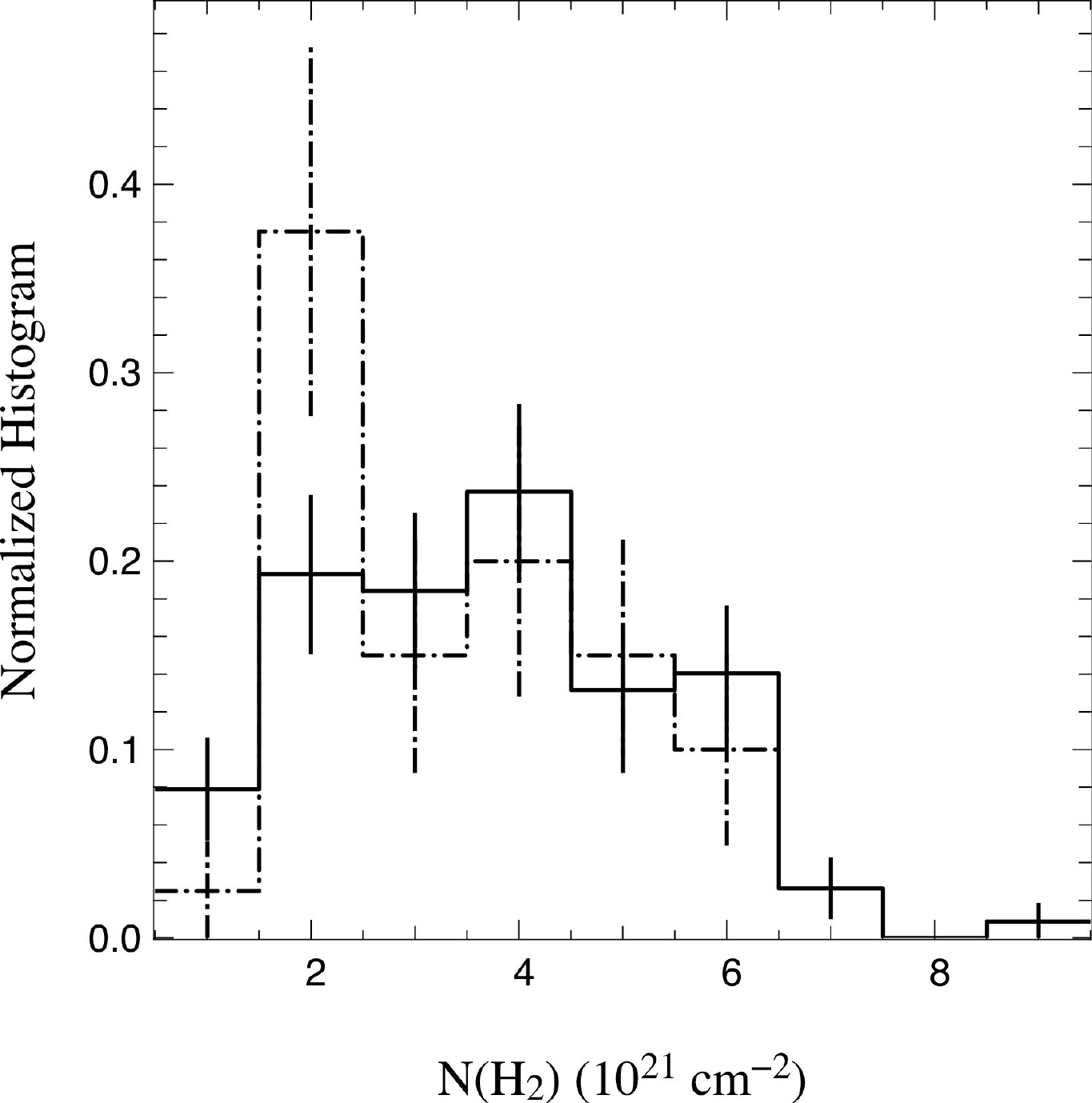}
\caption{N(H$_2$) histogram for BDs and TTS with $A_V<4$. Solid line : stars, dot-dashed line : BDs; the vertical lines indicate the  $\pm\,1\,\sigma$ error bars. \label{fig:bd*-nh2}}
\end{figure}

As the star density distribution is only a proxy for the actual density structure of the cloud where the objects were born, and 
in order to better track the relation between stars \& BDs and the underlying molecular gas, we have used the high spatial resolution molecular H$_2$ surface density map computed by Goldsmith et al. (2008) from recent $^{12}$CO and $^{13}$CO data, and we study the distribution of stellar and substellar objects relative to the actual cloud. We stress that strictly speaking,  the surface density ($\sigma$) is only a proxy for the volume density ($\rho$). However, a given object cannot be close to a volume density peak while far from a surface density peak. 
We measure the value of the H$_2$ column density at the position of every object,  star or BD, in the Taurus cloud, limiting ourselves to objects with $A_V<4$  as discussed above. Each measurement encompasses $\approx 3000\times 3000$\,AU at the Taurus distance. Figure~\ref{fig:bd*-nh2} shows the histogram of the underlying H$_2$ column density for stars (solid line) and BDs (dot-dashed line). The BD distribution shows a sharper peak at low surface densities, but  overall the two histograms are very similar.
We have performed a KS test on the two distributions and find that there is 36\% probability that the two samples are drawn from the same distribution. This result is consistent with previous studies of the nearest neighbor distance and shows that the distributions of both kinds of objects are the same with respect to the underlying molecular cloud. 


\subsection{Brown dwarfs with and without disks with respect to Taurus molecular gas} 
\label{sec:obj-dnd}
In this section, we use N(H$_2$), the H$_2$ surface density data to study the distribution of objects with and without disks relative to the molecular cloud. We apply the method to stars and BDs. Such a study is motivated by the fact that if TTS with and without disks appear relatively regularly distributed across the Taurus filament. However, there is a possibility that BDs with disks are more frequent in the northern part of the cloud (Guieu et al., 2007). 
If we split the Taurus filaments into two parts containing approximately the same number of objects (see dashed line in  figure~\ref{fig:fil-split}), and compute the disk fractions for TTS and BD in the northern and the southern filaments, we find that, while the TTS disk fraction is almost identical in both regions, the northern BD disk fraction is significantly higher than the southern one (see Table~\ref{tab:disk-frac}). 

\begin{figure}[htb]
\includegraphics[width=\hsize]{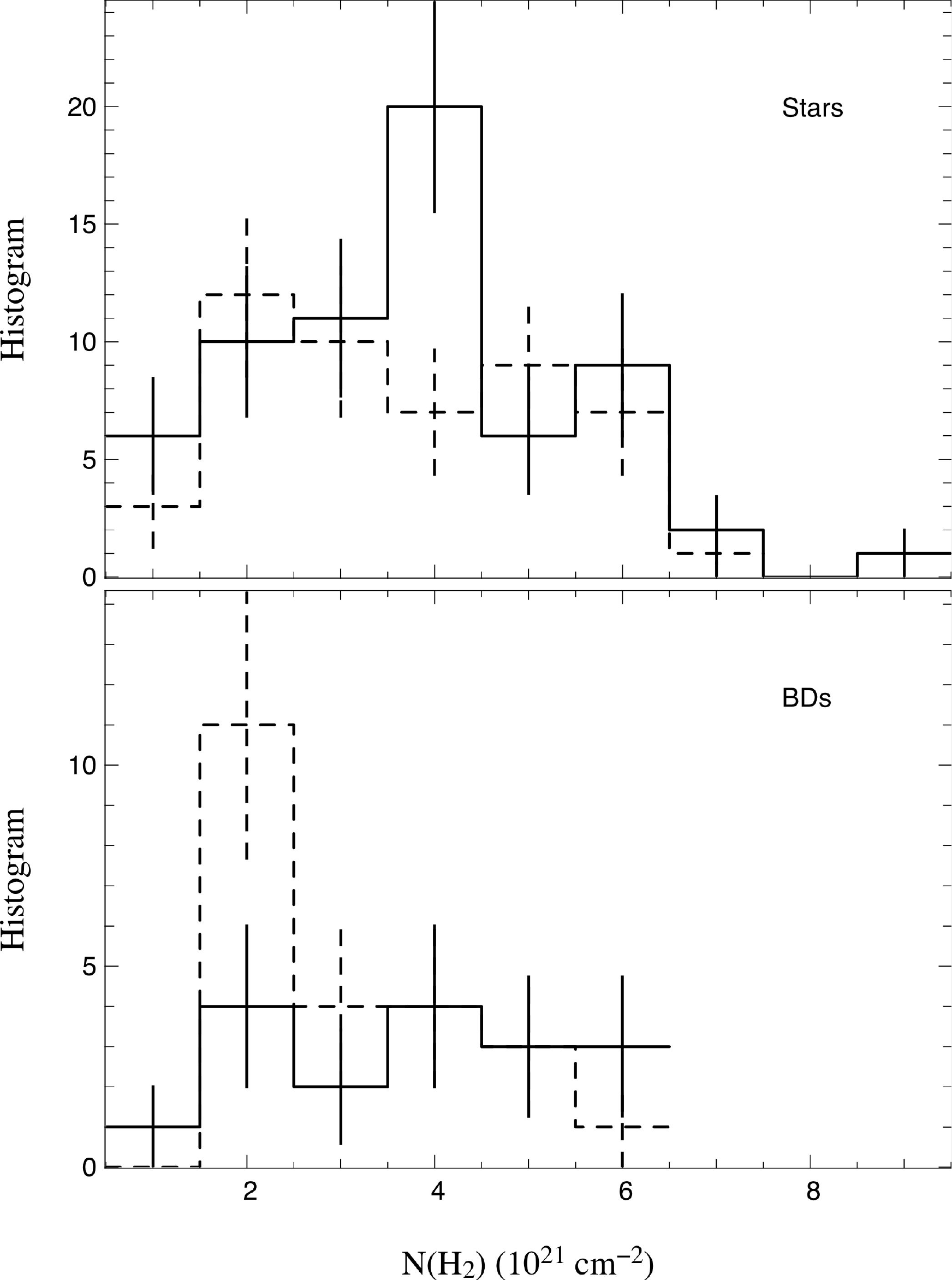}
\caption{Molecular hydrogen surface density distribution for stars (upper panel) and BDs (lower panel); in each plot, the solid line is the distribution of object with disk, and the dashed line, without disk.\label{fig:dist-nh2-sbd}}
\end{figure}

\begin{figure}[htb]
\includegraphics[width=\hsize]{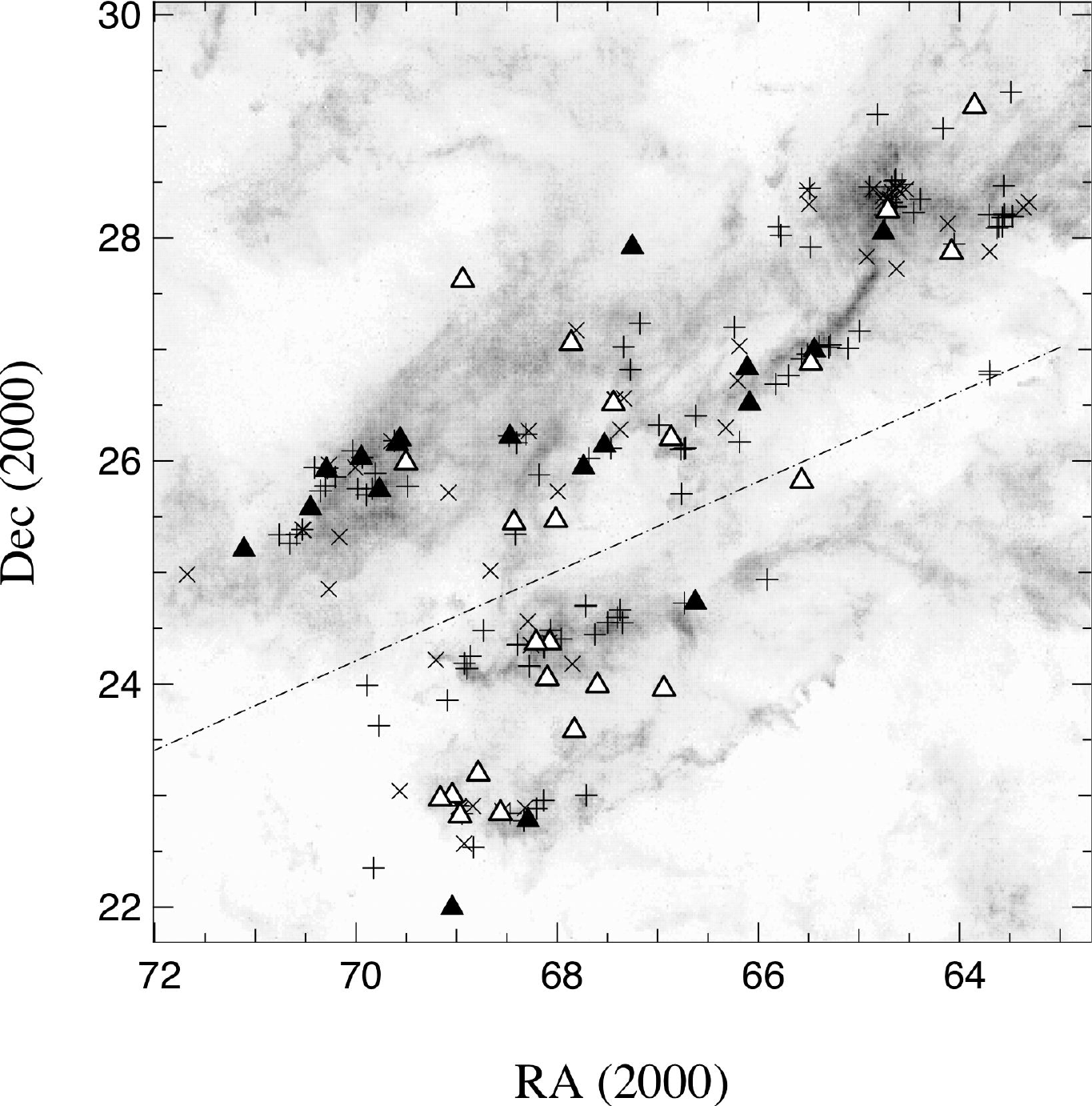}
\caption{Taurus TTS and BDs superimposed on CO molecular gas. The dashed line delineates the separation chosen between the northern and southern filaments (see text). Symbols are identical to those in figure~\ref{fig:colcol}.\label{fig:fil-split}}
\end{figure}

As in section~\ref{sec:*bd/gas}, we compute the underlying H$_2$ column density at the position of every object in the cloud (stars and BDs), with and without disks. The corresponding distribution of N(H$_2$) for BDs (with and without disk) and stars (with and without disk) is shown in figure~\ref{fig:dist-nh2-sbd}.
We then perform a KS test to estimate the probability that for the two kinds of objects (stars + BDs), both populations (with and without disks) are drawn from the same distribution. 
We find that the probability that each kind of object (BDs or stars), with and without disks are drawn from the same distribution is very high (18\%  for BDs, 31\% for stars). 
The distribution of objects with or without disks appears independent of the underlying gas distribution, suggesting  that the surrounding molecular cloud physical properties (at least density) are not directly linked to the existence of a disk around a newly formed object. 
Thus if the peculiar BD distribution mentioned by Guieu et al (2007) is real, it is not linked to the underlying cloud gas density. 
In a study on WTTS and CTTS in Taurus, Bertout et al. (2007) conclude that the disk lifetime is $4\times 10^6\,(M_*/M_\odot)^{0.75}\,$yr in this cloud. A variation of the disk lifetime with the central object' mass associated with small age difference between the filaments would be a possible explanation for the disk fraction difference between the northern and the southern part of the Taurus cloud. However, a shorter disk lifetime for BDs is very difficult to reconcile with BD disk fraction measurements in older star forming regions like Cha~I, IC\,348 and $\sigma$\,Ori (Luhman et al. 2005; 2008a,b) where it is similar to that of stars.
One possible explanation for the different disk fractions among stars
and BDs between the north and south (if it is real) is that
the southern clouds are indeed older, but the disk fraction for stars is
artificially high due to another explanation.
Surveys of Taurus have been biased toward finding
stars with disks rather than without disks. Many of the previous wide-field surveys have selected members based on disk
indicators like $H_\alpha$ and IR excess emission, whereas objects without disks 
are found by proper motions and the all-sky ROSAT survey, which are
rather shallow. In comparison, the BDs have been identified mostly
through optical color-magnitude diagrams, which are unbiased in terms of
disk presence. 


\begin{table}[htb]
\caption {CTTS and BD disk fraction in the northern and southern Taurus filament for objects with $A_V<4$.\label{tab:disk-frac}}
\begin{tabular}{lll} \hline\hline
  Filament  & TTS & BD \\
North   &  $37/68 = 54 \pm 9\,$\%  &  $14/25 = 56\pm 19\,$\% \\
South   &   $23/35 = 66\pm 14\,$\% & $3/14 = 21\pm 14\,$\% \\
 \hline
   \hline
\end{tabular}
\end{table}

 Testing the distribution of objects in the cloud relative to the gas was motivated by the fact that previous papers like Goodwin et al. (2005) obtained similar nearest neighbor distributions, even with ejections. 
The main result from our comparison of the BD and star distributions with respect to the underlying molecular gas is that BDs and stars are most probably drawn from the same distribution (p value of the null hypothesis is 0.36). This work complements and yields a picture consistent with the one published previously by Luhman (2006) in which BD and stars are distributed in a similar manner across the Taurus cloud, and present similar nearest neighbor distributions. 
The presence of $\approx$  two $\times$ more  BD than stars  in the first $2\times10^{21}\,$cm$^{-2}$ H$_2$ density bin remains to be explained. Yet, as both distributions appears to be statistically identical, this might be due to small number statistics effect. 

 \section{Conclusions}
\label{sec:concl}
We have presented new photometric measurements in the $3-8\,\mu$m domain from {\em Spitzer} IRAC  instrument, complementing previous IR photometry by Guieu et al. (2007) for 
more than 40 BDs in the Taurus cloud. We associate these IR measurements with new H$_2$ surface density computations from Goldsmith et al. (2008) to study and compare the distribution of stellar and substellar objects relative to the underlying molecular gas. Based on this larger sample, we update the disk fraction for stars and BDs in the Taurus cloud and find that the disk fraction around BDs is similar to the disk fraction in  their TTS counterparts.

We find that transition disks are very rare among BDs, a common result with TTS showing that the disk's disappearance timescale is also very short for substellar objects. These results call for further theoretical models of dust coagulation and settling as a function of the central object's mass. We model the only transition disk we find around CFHT~BD~Tau~12, an object that could have experienced more dust evolution than other members in Taurus. Note that Luhman et al (2006) find a slightly earlier spectral type than us for this object (M6 instead of M6.5) so CFHT~BD~Tau~12 could be an object at the border between VLMS and BD rather than a bona fide young BD.

We study the distribution of stars and BDs relative to the H$_2$ molecular gas surface density computed from CO observations and find that both distributions are likely drawn from the same distribution, even if there appears to be an excess of BDs found at low H$_2$ surface density.
A similar study for objects with and without disks shows that they have the same distribution relative to the underlying gas. 

Finally, the finding that the fact that BD with disks are  less numerous in the southern than in the northern portion of the Taurus filaments
could be a reflection of an older age for the southern part of the
cloud combined with a shorter lifetime for BD disks. A further
determination of the ages of the Taurus filaments via HR diagram for
instance is difficult due to the large dispersion one obtains in the
ages. 
Disk fraction measurements in Cha~I or IC 348 show no
evidence that the lifetimes of substellar objects are shorter.


 \begin{acknowledgements}
 We thank an anonymous referee for a very detailed and precise report that helped uncover several errors in the first version of this paper. 
  This research
has made use of the CDS database. We thank the Programme National
de Physique Stellaire (PNPS, CNRS/INSU, France) for financial support.
The authors also wish to extend special
thanks to those of Hawaiian ancestry on whose sacred mountain of Mauna Kea
we are privileged to be guests. Without their generous hospitality, the
CFH Telescope observations presented therein would not have been
possible. 
\end{acknowledgements}

\section{Bibliography}

Allard, F., Hauschildt, P.H,  \& Schweitzer, A. 2000, ApJ, 539, 366\\
Baraffe, I., Chabrier, G., Allard, F., \& Hauschildt, P. H. 1998, A\&A, 337, 403\\
Bate, M. R., Bonnell, I. A., \& Bromm, V. 2003, MNRAS, 339, 577\\
Bate, M. R. 2009, MNRAS, 392, 590\\
Bertout, C., Siess, L., \& Cabrit, S. 2007, A\&A, 473, L21\\
Bouy, H., Hu\'elamo, N., Pinte, C. et al. 2008, A\&A 486, 877\\
Chabrier, G., Baraffe, I., Allard, F., \& Hauschildt, P. 2000, ApJ, 542, 464\\
Damjanov, I., Jayawardhana, R., Scholz, A., et al. 2007, ApJ, 670, 1337\\
Fazio, G. G., Ashby, M. L. N., Barmby, P.  et al. 2004, ApJS, 154, 10\\
Goldsmith, P.F., Heyer, M., Narayanan, G., Snell, R., Li, D., \& Brunt, C. 2008, ApJ, 680, 428\\
Goodwin, S. P., Hubber, D. A., Moraux, E., Whitworth, A. P. 2005, AN, 326, 1040\\
Guieu, S., Dougados, C., Monin, J.-L., Magnier, E., Martin, E. L. 2006, A\&A, 446, 485\\
Guieu, S., Pinte, C., Monin, J.-L., et al.  2007, A\&A, 465, 855\\
Gutermuth, R. A., Myers, P. C., Megeath, S. T., et al.  2008, ApJ, 674, 336\\
Lay, O.P., Carlstrom, J.E., \& Hills, R.E., 1997, ApJ, 489, 917\\
Luhman, K. L., Stauffer, John R., Muench, A. A., Rieke, G. H., Lada, E. A., Bouvier, J., Lada, C. J. 2003, ApJ, 593, 1093\\
Luhman, K.  2004, ApJ, 617, 1216\\
Luhman, K. L., Lada, C. J., Hartmann, L., et al. 2005a, ApJ, 631, L69\\
Luhman, K. L., Adame, L.,  D'Alessio, P., Calvet, N., Hartmann, L., Megeath, S. T., Fazio, G. G. 2005b, ApJ, 635, L93\\
Luhman, K. L., Whitney, B. A., Meade, M. R., Babler, B. L., Indebetouw, R., Bracker, S., Churchwell, E. B. 2006, ApJ, 647, 1180\\
Luhman, K. L., Adame, L., D'Alessio, P., et al. 2007, ApJ, 666, 1219 \\
Luhman, K. L., Joergens, V., Lada, C., Muzerolle, J., Pascucci, I., White, R. 2007, Protostars \& Planets, V\\
Luhman, K. L., Allen,  L. E., Allen, P. R., et al. 2008a, ApJ, 675, 1375\\
Luhman, K. L., Hernandez, J., Downes, J. J., Hartmann, L., Brice–o, C. 2008b, ApJ, 688, 362\\
Mathis, J. S., \& Whiffen, G. 1989, ApJ, 341, 808\\
Padgett, D., Dougados, C.,  \& Guedel, M. 2007, Protostars \& Planets, V \\
Padgett, D. L., et al. 2008, Taurus Legacy Deli\-ve\-ry
Doc\-ument, \\available from
http://ssc.spitzer.caltech.edu/legacy/taurushistory.html \\
Padoan, P. \& Nordlund, \AA~2002, ApJ, 576, 870\\
Padoan, P. \& Nordlund, \AA~2004, ApJ, 617, 559\\
Pinte, C., M\'enard, F., Duch{\^e}ne, G., \& Bastien, P., 2006, A\&A, 459, 797\\
Pinte, C., Fouchet, L., M{\'e}nard, F., Gonzalez, J.-F., \& Duch{\^e}ne, G. 2007, A\&A, 469, 963\\
Pinte, C., Padgett, D. L., M\'enard, F., et~al. 2008, A\&A, 489, 633\\
Press, W. H., Teukolsky, S. A., Vetterling, W. T., \& Flannery, B. P. 1992, Numerical Recipes, The art of scientific computing, Cambridge University press, 1992, 2nd ed.\\
Reach, W. T., Megeath, S. T., Cohen, M., et al. 2005, PASP, 117, 978\\
Reipurth, B. \& Clarke, C. 2001, AJ, 122, 432\\
Scholz, A., Jayawardhana, R., \& Wood, K. 2006, ApJ, 645, 1498\\
Scholz, A., Jayawardhana, R., Wood, K., Meeus, G., Stelzer, B., Walker, C., O'Sullivan, M. 2007, ApJ, 660, 1517\\
Scholz, A. \& Jayawardhana, R. 2008, ApJ, 672, L49\\
Thies, I., \& Kroupa, P. 2007, ApJ, 671, 767\\ 
Werner, M. W.,  Uchida, K. I., Sellgren, K., Marengo, M., Gordon, K. D., Morris, P. W.,  Houck, J. R.,  Stansberry, J. A. 2004, ApJS, 154, 309\\
Whyatt, M. C. 2008, ARA\&A, 46, 339


\end{document}